\documentclass{elsart}
\usepackage[numbers,sort&compress]{natbib}
\usepackage{color,setspace,times}
\usepackage{amssymb,natbib,amsmath,graphicx,color,rotating,subfigure,url}
\usepackage{lineno}

\journal{Physica A}

\begin{document}

\begin{frontmatter}

\title{Memory effect and multifractality of cross-correlations in financial markets}
\author[SE]{Tian Qiu \corauthref{cor}},
\corauth[cor]{Corresponding author. Address: 696 South Fenghe
Avenue, School of Information Engineering, Nanchang Hangkong
University, Nanchang, 330063, China.} \ead{tianqiu.edu@gmail.com}
\author[SE]{Guang Chen},
\author[SA]{Li-Xin Zhong},
\author[SB]{Xiao-Wei Lei}

\address[SE]{School of Information Engineering, Nanchang Hangkong University, Nanchang, 330063, China}
\address[SA]{School of Journalism, Hangzhou Dianzi University, Hangzhou, 310018, China}
\address[SB]{Department of Physics, Chongqing University of Arts and Sciences, Chongqing 402160, China}

\begin{abstract}
Abstract: An average instantaneous cross-correlation function is
introduced to quantify the interaction of the financial market of a
specific time. Based on the daily data of the American and Chinese
stock markets, memory effect of the average instantaneous
cross-correlations is investigated over different price return time
intervals. Long-range time-correlations are revealed, and are found
to persist up to a month-order magnitude of the price return time
interval. Multifractal nature is investigated by a multifractal
detrended fluctuation analysis.

\end{abstract}


\end{frontmatter}

\section{Introduction}


In recent years, dynamics of financial markets has drawn much
attention of physicists
\cite{man95,man96,gha96,sta99,lal99,roe00,roe01,bou01,
sta01,gop99,lux99,gia01,qiu06,qiu08,qiu09,ren08,she09,she09a,tot06}.
Financial market is a complex system with many interacting
components. From the view of many-body systems, interactions among
components may lead the system to collective behavior, and therefore
result in the so-called dynamic scaling behavior. Based on large
amounts of historical data, some stylized facts have been revealed
in the past years, such as the 'fat tail' distribution of the price
return, and the long-range time-correlation of the magnitude of
returns \cite{man95,gop99,lux99,gia01}. Different models and
theoretical approaches have been developed to describe financial
markets\cite{lux99,gia01,cha97,sta99,con00,egu00,ren06,ren06a}.

The cross-correlation function is an important indicator to quantify
the interaction between stocks, and therefore has attracted much
attention of physicists in recent years. Random and nonrandom
properties of the cross-correlation and the relevant economic
sectors are revealed
\cite{ple02,cor05,she09,uts04,gar08,lal99,ple99,ple00,ple01,ros03,con09}.
Correlation-based hierarchical or network structures are studied
with the graph or complexity theory
\cite{man99,bon03,tum05,tum07,eom07,ery09,qiu10}. The so-called pull
effect is found with a time-dependent cross-correlation function
\cite{kul02}. These lines of work are mainly based on a static
definition of the cross-correlation function. The equal-time or the
time-dependent cross-correlation is usually defined as $C_{ij} =
{<r_{i}(t')r_{j}(t')>}$ or $C_{ij} = {<r_{i}(t')r_{j}(t'+\tau)>}$,
with $r_{i}(t')=lny_{i}(t'+\Delta t')-lny_{i}(t')$ being the return
of stock $i's$ price $y_{i}$ over a time interval $\Delta t'$,
$\tau$ being the time lag, and $<...>$ taking time average over
$t'$. The static cross-correlation function can not reveal the
dynamic behavior of the cross-correlations between stocks. More
recently, a Detrended Cross-Correlation Analysis(DCCA) is proposed
to investigate the memory effect of the cross-correlations between
two time series \cite{pod08a,pod08b,pod09a,pod09b}. Long-range
time-correlation of the cross-correlations is characterized by a
power-law scaling of the DCCA function. The DCCA method concentrates
on dynamics of two series' cross-correlations.

In this paper, we introduce an Instantaneous Cross-correlation($IC$)
and an Average Cross-correlation($AIC$) function by considering the
cross-correlations of a single time step. The $IC$ and $AIC$
function describes the current interaction between stocks with local
information. Our purpose is to investigate the dynamics of the $IC$
and $AIC$ series, based on the daily data of the American and
Chinese stock markets. More importantly, we examine the memory
effect of the $AIC$ over different price return time intervals. The
multifractal nature of the $AIC$ is also revealed.

The organization of this paper is as follows. In the next section,
the datasets and the definition of the $IC$ and $AIC$ functions are
presented. In Sec. 3, we investigate the memory effect of the $IC$
and $AIC$ for a shorter price return time interval. In Sec. 4, the
memory effect of the $AIC$ is detected over different scales of
price return time intervals. In section 5, we examine the
multifractal nature of the $AIC$. Finally, Sec. 6 contains the
conclusion.

\section{datasets and instantaneous cross-correlations}

To obtain a comprehensive study, we analyze two different databases,
the New York Stock Exchange (NYSE) and the Chinese Stock
Market(CSM). The two markets cover the mature and the emerging
markets. The NYSE is one of the oldest stock exchanges, whereas the
CMS is a newly set up market in $1990$. We investigate the daily
data of $249$ individual stocks, with $2900$ data points from the
year $1997$ to $2008$ for the NYSE, and the daily data of $259$
individual stocks, with $2633$ data points from the year $1997$ to
$2007$ for the CSM. To compare different stocks, we define the
normalized the price return as
\begin{equation}
R_{i}(t',\Delta t')={\frac {r_{i}(t')-<r_{i}(t')>} {\sigma_{i}}}
\label{e10}
\end{equation}

where $r_{i}(t')$ is the price return of stock $i$ at time $t'$, and
$\Delta{t'}$ is the price return time interval. The
$\sigma_{i}={\sqrt{<r_{i}^{2}>-<r_{i}>^{2}}}$ is the standard
deviation of $r_{i}$, and $<\ldots>$ takes time average over $t'$.
In order to quantify the current cross-correlation between stocks,
we introduce an $IC$ function between two stocks by

\begin{equation}
IC_{ij}(t')=R_{i}(t')R_{j}(t'), \label{e20}
\end{equation}

The $IC$ function indicates the instantaneous cross-correlation
between two individual stocks. However, it does not depict the
average interaction of the financial market. Therefore, we define an
$AIC$ function as

\begin{equation}
AIC(t')=\frac{2}{N(N-1)}\sum _{i=1}^{N-1} \sum _{j=i+1}^{N}
C_{ij}(t') \label{e40}
\end{equation}

where $N$ is the number of stocks. The $AIC$ function indicates the
average instantaneous cross-correlation of a number of stocks with
the stock size to be $N$. As the stock number $N$ is large enough,
the $AIC$ function can be then considered as an indicator to
quantify the average interaction of the financial market at a
specific time step. For $N=2$, the $AIC$ function reduces to the
$IC$ function.

\section{Memory effect of $IC$ and $AIC$ for a shorter $\Delta t'$}

It is important to measure the memory effect of the time series
during the dynamic evolution. We investigate the memory effect of
the $IC$ and $AIC$ by computing the time-correlations. The
autocorrelation function is widely adopted to measure the
time-correlation. However, it shows large fluctuations for
nonstationary time series. Therefore, we apply the DFA method \cite
{pen94,pen95}.

Considering a fluctuating dynamic series $A(t')$, one can
construct
\begin{equation}
B(t')=\sum\limits_{t"=1}^{t'} A(t"), \label{e8}
\end{equation}
Dividing the total time interval into windows $N_t$ with a size of
$t$, and linearly fit $B(t')$ to a linear function $B_{t}(t')$ in
each window. The DFA function of the $k_{th}$ window box is then
defined as:
\begin{equation}
f_{k}(t)^{2}=\frac{1}{t}\sum\limits_{t'=(k-1)t+1}^{kt}{[B(t')-B_t(t')]}^2
, \label{e9}
\end{equation}

The overall detrended fluctuation is estimated as
\begin{equation}
F_{2}(t)^{2}=\frac{1}{N_t}\sum\limits_{k=1}^{N_t}{[f_k(t)]}^2 ,
\label{e9}
\end{equation}

In general, $ F_{2}(t)$ will increase with the window size $t$ and
obey a power-law behavior $F_{2}(t)\sim t^{H}$. If $0.5<H<1.0$,
$A(t')$ is long-range correlated in time; if $0<H<0.5$, $A(t')$ is
temporally anti-correlated; $H=0.5$ corresponds to the Gaussian
white noise, while $H=1.0$ indicates the $1/f$ noise. If $H$ is
bigger than $1.0$, the time series is considered to be unstable.

The DFA functions of the $IC$ and the $AIC$ are computed with the
price return time interval $\Delta t' = 1$ day. To illustrate the
results, we take $6$ $ICs$ and $AICs$ as examples, with the stocks
randomly chosen from the NYSE and the CSM, respectively. As shown in
Fig. 1(a), the DFA exponents of the $ICs$ are estimated to be from
$0.46$ to $0.65$ for the NYSE. The exponent $0.46$ is close to the
Gaussian behavior, while $0.65$ is the long-range correlation.
Similarly, as shown in Fig. 1(c) for the $ICs$ of the CSM, the DFA
exponents range from $0.53$ to $0.67$, also corresponding to the
Gaussian behavior and the long-range correlation, respectively. It
suggests that the long-range time-correlation does not hold for all
$ICs$. However, when we compute the DFA of the $AIC$, robust
long-range time-correlations are observed for both the NYSE and the
CSM. As shown in Fig. 1(b) and (d), the DFA functions of 6 $AICs$
are shown as examples with $N=50$. The DFA exponents are estimated
to be around 0.73 for the NYSE, and 0.67 for the CSM, both in the
long-range time-correlation range $(0.5,1.0)$. The DFA exponents of
the $AIC$ take the similar value for a larger stock number $N$ from
our databases. It implies that, for both the NYSE and the CSM, even
though the absence of the long-range time-correlation of the
instantaneous cross-correlation between two individual stocks, the
average instantaneous cross-correlation of a number of stocks is
long-range correlated, i.e., the average interaction of the
financial market shows long-term memory. The result is reasonable.
For example, it is possible for two correlated companies to break up
their relationship during the time evolution for some reason. With
the end of the correlation, the memory of the cross-correlation then
also ends up. However, the fluctuation from the endogenous events
would not influence the average interaction of the whole market,
i.e., as a collective, the cross-correlation of the financial market
is always characterized by a long-range memory.

\section{Memory effect of $AIC$ for different $\Delta t'$}

Scalings observed in the financial market has been found to always
evolve with the different value of the price return intervals. For
example, the 'fat tail' of the probability density function of the
price returns can not be found for a big return time interval \cite
{gop99}. To further understand the memory effect of the
cross-correlations, we then investigate the DFA functions of the
$AIC$ with different price return time interval $\Delta t'$. The
return time interval $\Delta t'$ covers three magnitude orders, the
day, the week and the month time scales.

The $AIC$ is computed with $N=249$ for the NYSE, and $N=259$ for the
CSM. In Fig. 2(a), the DFA function of the $AIC$ is shown for the
NYSE, with $\Delta t' = 1$, $5$, $10$, $22$, and $44$ days,
approximate to a working day, a week, half a month, a month, and two
months. For $\Delta t' = 1$ day, clean power-law behavior is
observed, with the exponent estimated to be $0.74$, consistent with
the exponents obtained in Fig. 1. For the return time interval
$\Delta t'$ bigger than $1$ day, two-stage power-law scalings are
observed, with a crossover in between. Such a two-stage behavior has
been widely found in the DFA function of the financial series, such
as the volatilities, intertrade durations, etc
\cite{liu99,qiu07,jia08a}. The crossover time is about $t_{c} \sim
35$ days. For the smaller window size, the DFA exponents take the
value around $1.0$ for $\Delta t'=5$, and bigger than the 1.0 for
$\Delta t'=10$, $22$ days, which correspond to the $1/f$ noise and
unstable time series, respectively. Due to the narrow range of the
smaller window size, we care more about the DFA exponents of the
larger window size. For the larger window size from $t=35$ to 100
days, the exponents are measured to be $0.65$, $0.72$, $0.86$ for
$\Delta t'=5$, $10$ and $22$ days, with all the exponent value
ranged in the long-range time-correlation. The estimated exponents
also remain unchanged for a relatively larger window size than 100
days. However, due to the finite size of the time series, it will
show large fluctuation for a large window size. For $\Delta t'=44$
days, both the smaller and the larger window size do not show
long-range time-correlations. Therefore, the long-range
time-correlation of the $AIC$ persists up to a working month
magnitude of the price return time interval for the NYSE for the
large window size. Similar behavior is also observed for the CSM.
For $\Delta t' = 1$ day, clean power-law behavior is observed, with
the exponent estimated to be 0.68, around 0.67 found in Fig. 1.
Also, two-stage power-law scalings are observed for $\Delta t'=5$,
$10$ days, with the smaller window size showing $1/f$ noise and
unstable time series, and the larger window size showing long-range
time-correlations. The cross-over time $t_{c} \sim 22$ days. The
long-range memory persists up to half a month magnitude of the price
return time interval for the CSM for the larger window size.

\section{multifractal nature of $AIC$}

Financial time series such as the price returns and the intertrade
durations has been revealed to present multifractal feature
\cite{jia08a,jia08b,zho09}. The multifractal detrended fluctuation
analysis(MF-DFA) has been successfully applied to detect
multifractal characteristic of nonstationary time series
\cite{kan02}. We then apply the MF-DFA into the $AIC$, with $N=249$
for the NYSE, and $N=259$ for the CSM. The MF-DFA is a
generalization of the DFA method by considering different order of
detrended fluctuation. For the $q_{th}$ order of the detrended
fluctuation, we have

\begin{equation}
F_{q}(t) =\{ \frac{1 }{N_t}\sum_{k=1}^{N_t}[f_{k}(t)]^q\}^{1/q},
\label{e9}
\end{equation}

where $q$ can take any real number except $q = 0$. For $q = 0$, we
have
\begin{equation}
F_{0}(t) =\exp\{ \frac{1 }{N_t}\sum_{k=1}^{N_t}\ln[f_{k}(t)]\},
\label{e9}
\end{equation}

The MF-DFA function $F_q(t)$ scales with the window size $t$:

\begin{equation}
F_{q}(t)\sim t^{h(q)}, \label{e9}
\end{equation}

where $h(q)$ is the MF-DFA exponent, with $q = 2$ recovering the DFA
exponent. Due to the finite size of the time series, the $F_q(t)$
shows large fluctuations for the large values of $|q|$. Here we take
$q \in [-2, 4]$, and $\Delta t'=1$ day as examples to show the
multifractal properties. The $F_q(t)$ of the $AIC$ is shown for the
NYSE and CSM in Fig. 3. Clean power law scalings are observed for
$q=-2$, $0$, $2$ and $4$, with the exponents estimated to be $1.13$,
$0.95$, $0.68$, $0.52$ for the NYSE, and $0.89$, $0.84$,
$0.74$,$0.57$ for the CSM. The dependence of the $F_q(t)$ on $q$
suggests that the $AIC$ shows a multifractal characteristic. The
MF-DFA exponent $h(q)$ versus different $q$ is shown in Fig. 4(a)
and (d), respectively for the NYSE and the CSM.

The scaling exponent function $\tau(q)$ based on partition function
is widely adopted to reveal the multifractality,

\begin{equation}
\tau(q)=qh(q)-D_{f}, \label{e9}
\end{equation}

where $D_f$ is the fractal dimension, with $D_f = 1$ in our case. As
shown in Fig. 4(b) and (e), the $\tau(q)$ of the NYSE and the CSM
presents a strong nonlinearity, which is consistent with
multifractal characteristic. By the Legendre transformation, the
local singularity exponent $\alpha$ and its spectrum $f(\alpha)$ can
be calculated as \cite{hal86},

\begin{equation}
\alpha=d\tau(q)/dq, \label{e9}
\end{equation}
\begin{equation}
f(\alpha)=q\alpha-\tau(q), \label{e9}
\end{equation}

The difference between the maximum and the minimum of the local
singularity exponent $\Delta\alpha \triangleq
\alpha_{max}-\alpha_{min}$ is widely used to quantify the width of
the extracted multifractal spectrum. The larger the $\Delta\alpha$,
the stronger the multifractality. Fig. 4(c) and Fig. 4(f) illustrate
the multifractal singularity spectra $f(\alpha)$, with the width of
the extracted multifractal spectrum $\Delta\alpha$ measured to be
$0.64$ and $0.93$ respectively for the NYSE and the CSM. It
indicates the CSM shows stronger multifractality than the NYSE.

\section{Conclusion}

We have investigated the memory effect of the instantaneous
cross-correlations and the average instantaneous cross-correlations
based on the daily data of the NYSE and the CSM. It is interesting
to find that, in spite of the absence of the long-range
time-correlation of the instantaneous cross-correlations between two
individual stocks, the average instantaneous cross-correlation of a
set of stocks is long-range correlated for the price return time
interval $\Delta t'=1$ day. The long-range time-correlation persists
up to a month price return time interval for the NYSE, and half a
month time interval for the CSM for the large time window.

Multifractal nature is revealed for the average instantaneous
cross-correlations by the MF-DFA. By examining the MF-DFA function
$F_q(t)$, the scaling exponent function $\tau(q)$, and the extracted
multifractal spectrum $f(\alpha)$, multifractal features are
revealed for both the NYSE and the CSM.

\bigskip
{\textbf{Acknowledgments:}}

This work was supported by the National Natural Science Foundation
of China (Grant Nos. 10805025, 10774080) and the Foundation of
Jiangxi Educational Committee of China.

\bibliography{rint}

\begin{figure}[htb]
\centering
\includegraphics[width=8.5cm]{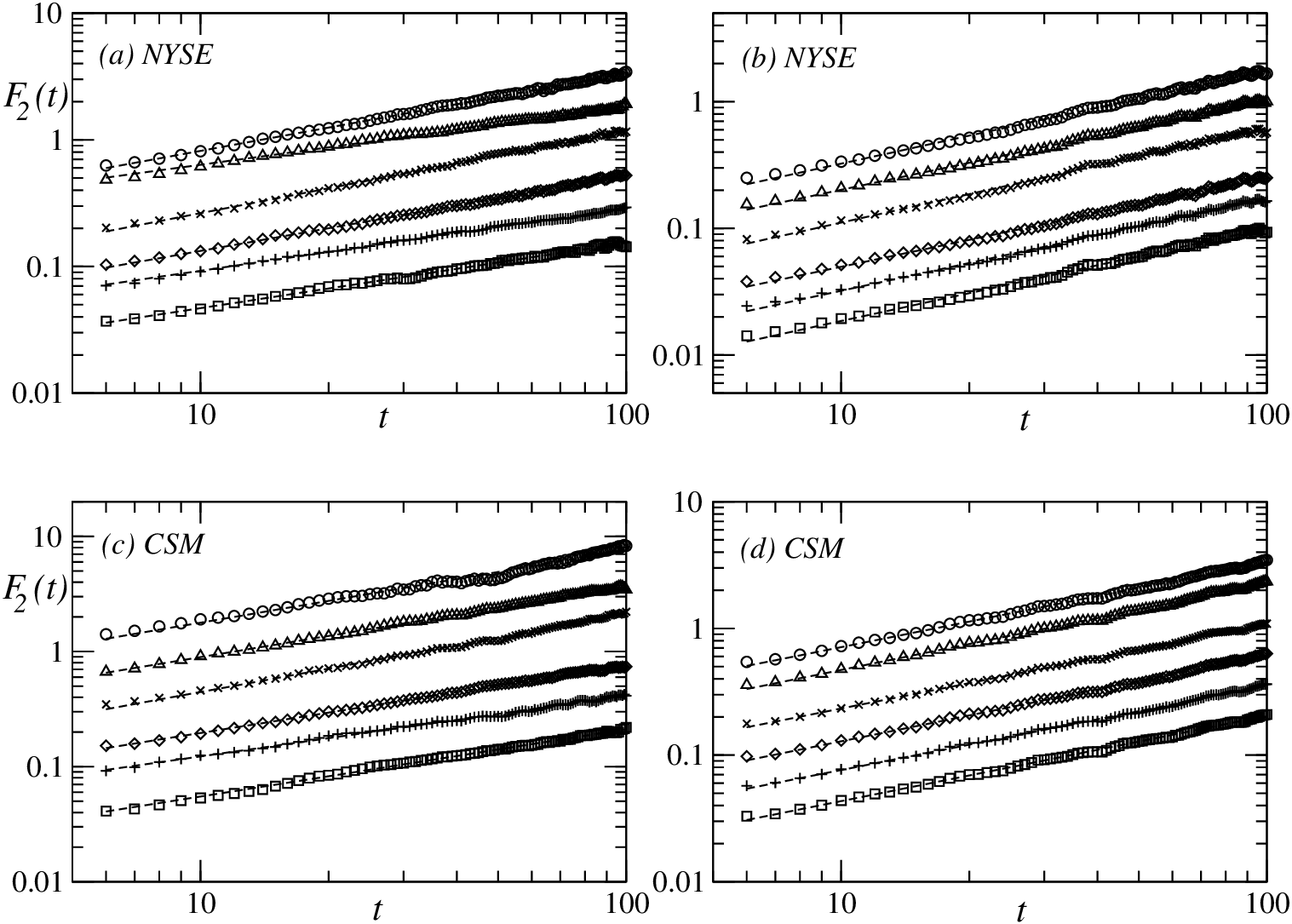}
\caption{\label{Fig:1} The DFA functions of the $IC$ and the $AIC$
are displayed on a log-log scale,  with the circles, triangles,
crosses, diamonds, pluses and squares being six samples. The dashed
lines are the power law fits. For clarity, some curves have been
shifted downwards or upwards.(a) for the $IC$ of the NYSE. The
exponents are measured to be $0.61$, $0.46$, $0.65$, $0.60$, $0.49$
and $0.52$. (b) for the $AIC$ of the NYSE with $N=50$. The exponents
are measured to be $0.73$, $0.71$, $0.75$, $0.71$, $0.73$ and
$0.73$. (c) for the $IC$ of the CSM. The exponents are measured to
be $0.61$, $0.67$, $0.63$, $0.59$, $0.53$ and $0.58$. (d) for the
$AIC$ of the CSM with $N=50$. The exponents are measured to be
$0.67$, $0.68$, $0.67$, $0.68$, $0.66$ and $0.67$. }
\end{figure}

\begin{figure}[htb]
\centering
\includegraphics[width=8.5cm]{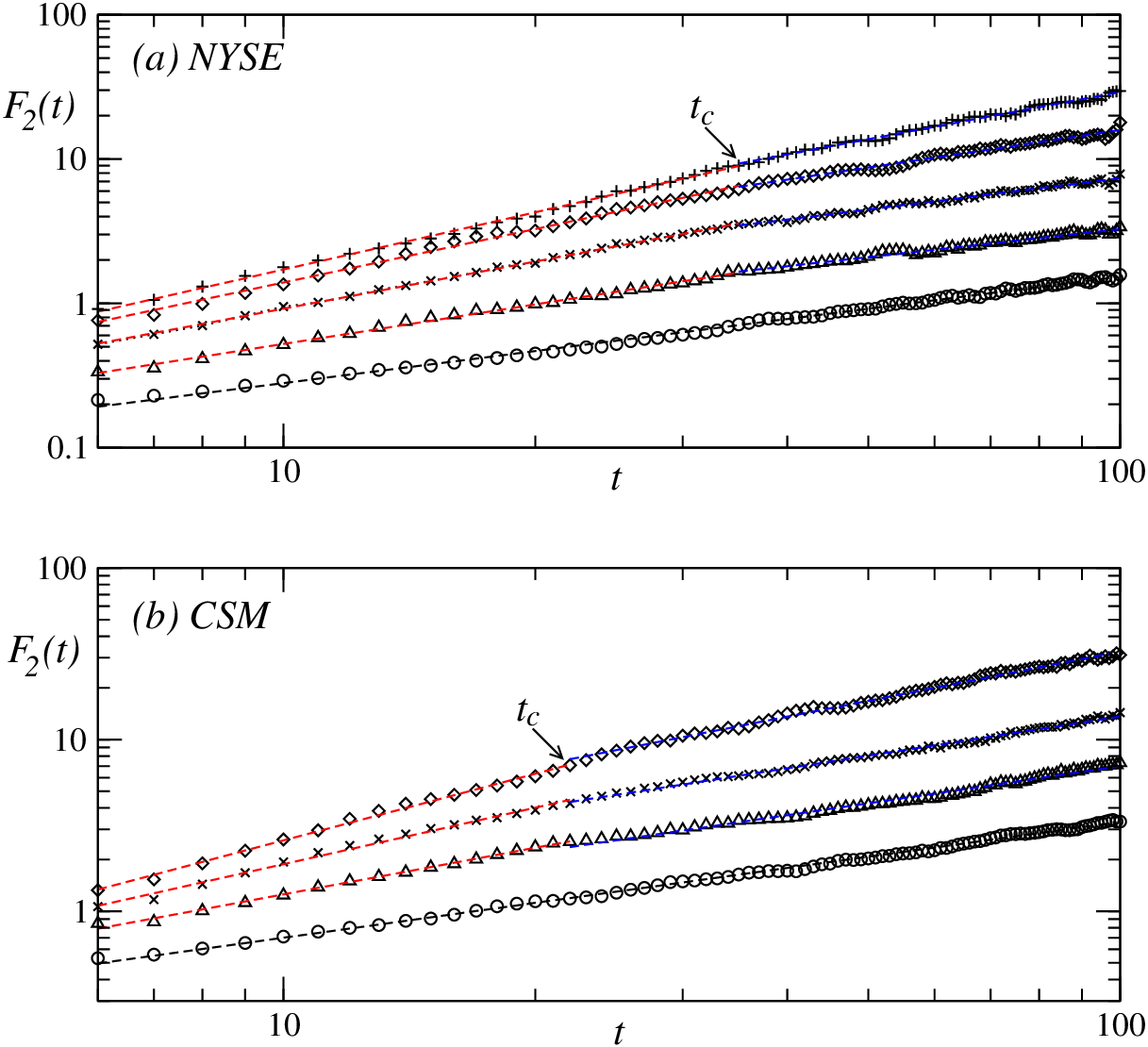}
\caption{\label{Fig:2} The DFA functions of the $AIC$ are displayed
on a log-log scale, with the dashed lines being the power law fits.
Some curves have also been shifted downwards or upwards for clarity.
(a) For the NYSE and $N=249$, the circles, triangles, crosses,
diamonds, and pluses are for $\Delta t'=1$, $5$, $10$, $22$, and
$44$ days, respectively. The power-law exponent for $\Delta t'=1$
day is measured to be $0.74$ for the whole window size. Two stage
power-law exponents for $\Delta t'=5$, $10$, $22$ and $44$ days are
respectively measured to be $0.91$, $1.08$, $1.23$, $1.31$ for the
smaller window size, and $0.65$, $0.72$, $0.86$, $1.08$ for the
larger window size. (b)For the CSM and $N=259$, the circles,
triangles, crosses, and diamonds are for $\Delta t'=1$, $5$, $10$,
and $22$ days, respectively. The power-law exponent for $\Delta
t'=1$ day is measured to be $0.68$ for the whole window size. Two
stage power-law exponents for $\Delta t'=5$, $10$ and $22$ days are
respectively measured to be $0.90$, $1.09$, $1.29$ for the smaller
window size, and $0.72$, $0.76$, $0.95$ for the larger window size.}
\end{figure}

\begin{figure}[htb]
\centering
\includegraphics[width=8.5cm]{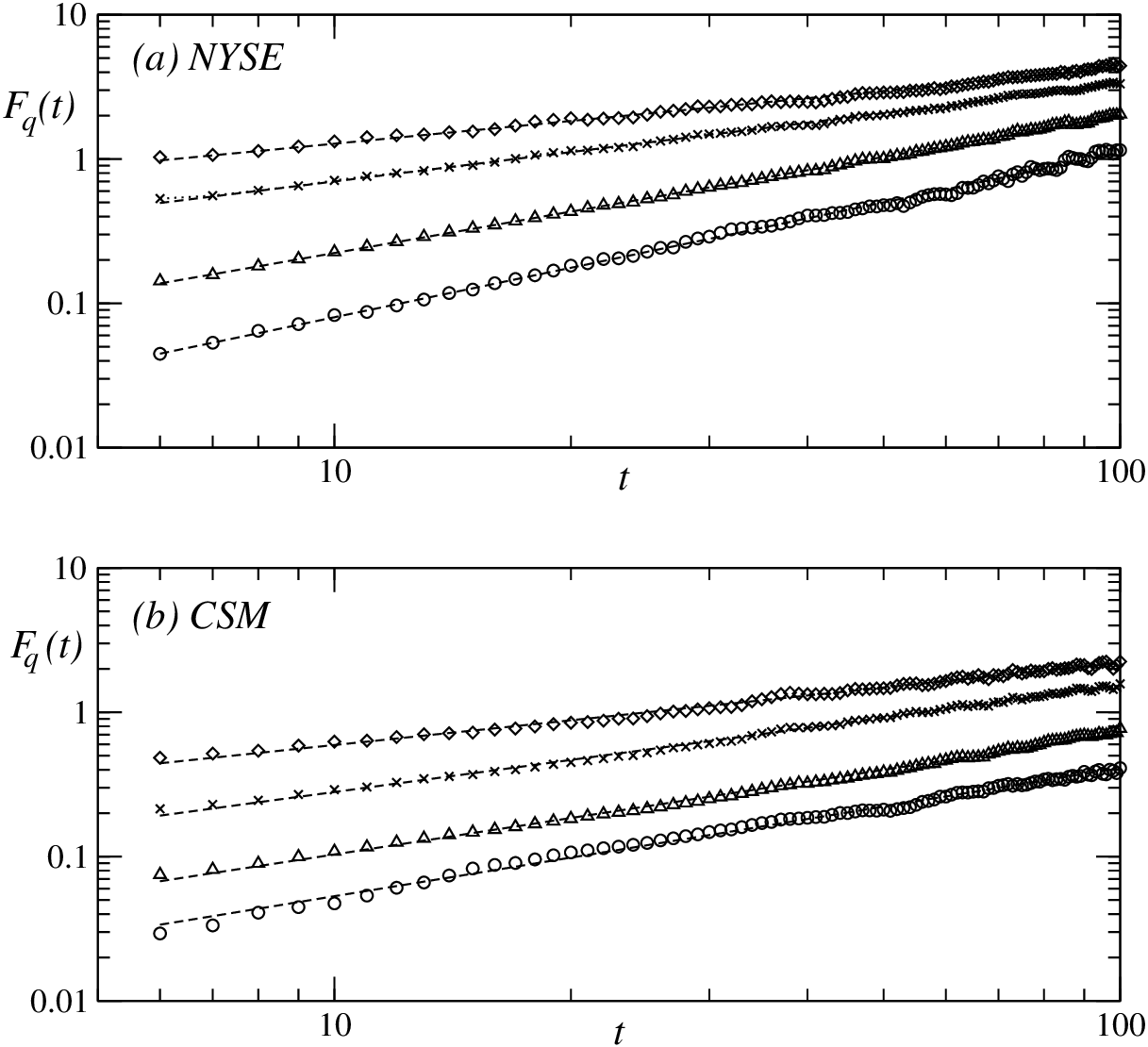}
\caption{\label{Fig:3} The MF-DFA functions of the $AIC$ are
displayed on a log-log scale, with the dashed lines being the power
law fits. The circles, triangles, crosses and diamonds are for
$q=-2$, $0$, $2$ and $4$, with curves being shifted downwards or
upwards for clarity. (a) For the NYSE and $N=249$, with the
exponents estimated to be $1.13$, $0.95$, $0.68$, and $0.52$. (b)
For the CSM and $N=259$, with the exponents estimated to be $0.89$,
$0.84$, $0.74$, and $0.57$.}
\end{figure}

\begin{figure}[htb]
\centering
\includegraphics[width=8.5cm]{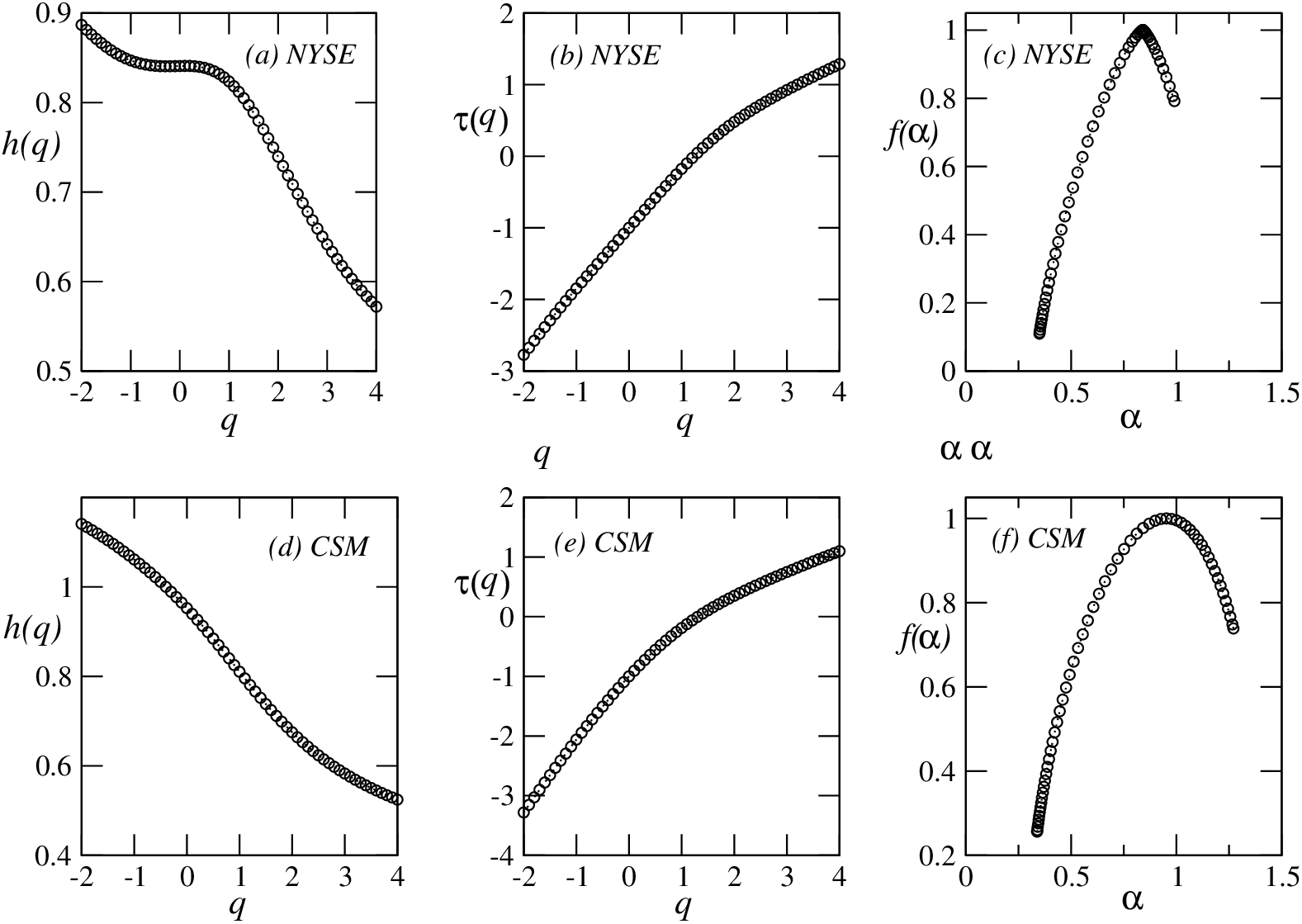}
\caption{\label{Fig:4} The multifractal analysis of the $AIC$ is
displayed with $N=249$ for the NYSE and  $N=259$ for the CSM. (a)
and (d) are the MF-DFA exponents $h(q)$ versus $q$ for the NYSE and
the CSM. (b) and (e) are the scaling exponent function $\tau(q)$ for
the NYSE and the CSM. (c) and (f) are the multifractal spectrum
$f(\alpha)$ for the NYSE and the CSM.}
\end{figure}

\end{document}